\documentclass{article}
\usepackage{spconf,amsmath,graphicx}

\usepackage{multirow}
\usepackage{graphicx}
\usepackage{amssymb, amsmath, bm, mathtools,array}
\usepackage{textcomp}
\usepackage{hyperref}
\usepackage{subcaption}
\usepackage{verbatim,lipsum}

\RequirePackage[normalem]{ulem} 
\RequirePackage{color}\definecolor{BLUE}{rgb}{0,0.20,0.75} 
\RequirePackage{color}\definecolor{BROWN}{RGB}{60,128,49} 

\title{Investigation of enhanced Tacotron text-to-speech synthesis systems with self-attention for pitch accent language}
\name{Yusuke Yasuda$^{1,3}$, Xin Wang$^1$, Shinji Takaki$^1$, Junichi Yamagishi$^{1,2}$\sthanks{This work was partially supported by JST CREST Grant Number JPMJCR18A6, Japan and by MEXT KAKENHI Grant Numbers (16H06302, 17H04687, 18H04120, 18H04112, 18KT0051), Japan.}}
\address{$^1$National Institute of Informatics, Japan
~~~$^2$The University of Edinburgh, Edinburgh, UK\\
$^3$SOKENDAI (The Graduate University for Advanced Studies), Japan\\
  {\small \tt yasuda@nii.ac.jp, wangxin@nii.ac.jp, takaki@nii.ac.jp,  jyamagis@nii.ac.jp}}

\begin{document}
\ninept
\maketitle
\begin{abstract}

End-to-end speech synthesis is a promising approach that directly converts raw text to speech. Although it was shown that Tacotron2 outperforms classical pipeline systems with regards to naturalness in English, its applicability to other languages is still unknown. Japanese could be one of the most difficult languages for which to achieve end-to-end speech synthesis, largely due to its character diversity and pitch accents. Therefore, state-of-the-art systems are still based on a traditional pipeline framework that requires a separate text analyzer and duration model.
Towards end-to-end Japanese speech synthesis, we extend Tacotron to systems with self-attention to capture long-term dependencies related to pitch accents and compare their audio quality with classical pipeline systems under various conditions to show their pros and cons. In a large-scale listening test, we investigated the impacts of the presence of accentual-type labels, the use of force or predicted alignments, and acoustic features used as local condition parameters of the Wavenet vocoder. Our results reveal that although the proposed systems still do not match the quality of a top-line pipeline system for Japanese, we show important stepping stones towards end-to-end Japanese speech synthesis.

\end{abstract}
\begin{keywords}
speech synthesis, deep learning, Tacotron
\end{keywords}

\section{Introduction}
\label{sec:intro}

Tacotron \cite{Wang2017} opened a novel path to end-to-end speech synthesis. It enables us to directly convert input text to audio. Unlike traditional pipeline methods that typically consist of separate text analyzer, acoustic, and duration models, Tacotron handles everything as a single model, which reduces laborious feature engineering and error propagation across cascaded models. Indeed, Tacotron2, which is a combination of the Tacotron system and WaveNet \cite{wavenet}, successfully generated audio signals that resulted in very high MOS scores comparable to human speech \cite{Shen2017}.

The above achievements of Tacotron and Tacotron2 and similar results reported for Clarinet \cite{Ping2018}, and Transformer based TTS \cite{DBLP:journals/corr/abs-1809-08895} are confirmed only for English, and there have been only a few investigations into such architectures with other languages to the best of our knowledge. This is partially or mainly because additional challenges must be overcome for other languages. This study focuses on the Japanese language, which is among the most challenging languages.

Japanese writing has three types of orthographical characters: Hiragana, Katakana, and Kanji (Chinese). The diversity of characters in Japanese causes a critical problem related to rare characters. Moreover, Japanese is a pitch-accented language, and accentual-types (accent nucleus positions) may change the meanings of words. However, accentual-types are not explicitly shown in Japanese characters. Moreover, due to the accent sandhi phenomena, accent nucleus positions are context dependent, so they change positions depending on adjacent words. Because of these problems, state-of-the-art systems for Japanese are dominantly pipeline systems that still rely on an external text analyzer including hand-written dictionaries and rules of pitch accent types for each word or word-to-accentual-type predictors trained on such external resources \cite{Bruguier2018}. An end-to-end approach may potentially simplify these process in data driven way.

Towards the development of end-to-end Japanese TTS systems, we apply the Tacotron system to the Japanese language. We first propose enhanced systems with self-attention to capture long-term dependency better. We then compare their audio quality with that of classical pipeline systems under various conditions. Finally, we conduct a large-scale listening test to investigate the impacts of the presence of accentual-type labels, the use of force- or predicted alignments, and acoustic features used as local condition parameters of the Wavenet vocoder. 

The remaining part of this paper is structured as follows. In Section 2, we describe our Japanese Tacotron systems enhanced with self-attention. Section 3 shows experimental conditions and the results of a large-scale listening test. Section 4 concludes with our findings and our future work. 

\section{Proposed architectures for Japanese TTS}
\vspace{-2mm}
\subsection{Tacotron using phoneme and accentual type}
\label{subsec:japanese}

\begin{figure}[tb]
\begin{center}
\includegraphics[width=0.98\columnwidth]{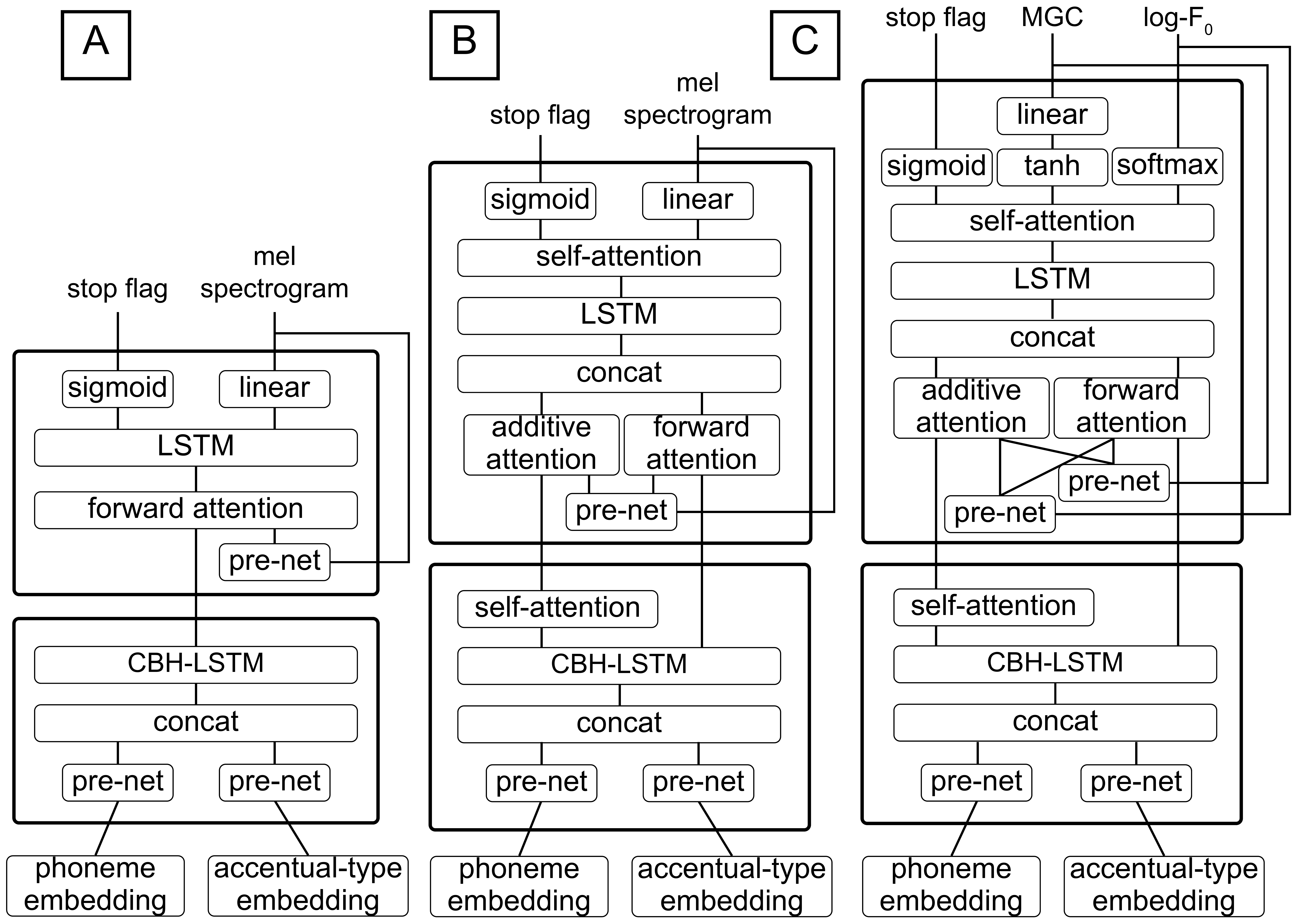}
\vspace{-18pt}     
\end{center}
\caption{Architectures of proposed systems with accentual-type embedding. A: \textit{JA-Tacotron}. B: \textit{SA-Tacotron}, C: \textit{SA-Tacotron} using vocoder parameters.}
\label{fig:tacotron}
\vspace{-20pt}   
\end{figure}

In this section, we describe our slightly modified baseline Tacotron \cite{Wang2017} that can handle Japanese accentual-type labels. We refer to this system as \textit{JA-Tacotron}. Figure \ref{fig:tacotron}-A shows its architecture. Tacotron is a sequence-to-sequence architecture \cite{sutskever2014sequence} that consists of encoder and decoder networks. Unlike classical pipeline systems with explicit duration models, Tacotron uses an attention mechanism \cite{Bahdanau2014} that implicitly learns alignments between source and target sequences. In this paper, we use phoneme and accentual-type sequences as a source and mel-spectrogram as a target as our first investigation towards end-to-end Japanese speech synthesis. This baseline architecture is inspired from \cite{Zhang2018}, which applied Tacotron to the Chinese language. On the encoder side, phoneme and accentual-type sequences are embedded to separate embedding tables with different dimensions, and the embedding vectors are bottle-necked by their corresponding pre-nets \cite{Wang2017}. The two inputs are then concatenated and encoded by Convolution Banks, Highway networks, bidirectional-LSTM (CBH-LSTM) with zoneout regularization \cite{Krueger2016}.

At the decoder, encoded values are decoded with attention based LSTM decoder. We use forward attention \cite{Zhang2018} instead of additive attention \cite{Bahdanau2014} as an attention mechanism. As suggested in \cite{Zhang2018}, the forward attention accelerates the alignment learning speed and provides distinct and robust alignment with less training time than the original Tacotron. The decoder LSTM is regularized with zoneout as well as the encoder since it is expected that the zoneout regularization will reduce alignment errors. We set the reduction factor to be two so that the decoder outputs two frames at each time step. A predicted mel-spectrogram is converted to an audio waveform with WaveNet \cite{wavenet}. We use a frame shift of 12.5 ms for the mel-spectrogram to train the \textit{JA-Tacotron} model as in \cite{Shen2017}\footnote{Our WaveNet model for \textit{JA-Tacotron} is trained by fine-tuning using a ground truth mel-spectrogram with a frame shift of 12.5~ms starting with an existing model trained with a mel-spectrogram with a frame shift of 5~ms in order to make comparison with TTS systems using vocoder parameters fairer. We use softmax distribution as an output layer of WaveNet.}

\vspace{-2mm}
\subsection{Extending Tacotron with self-attention}
\label{subsec:extended_system_with_self_attention}

A pitch-accent language like Japanese uses lexical pitch accents that involve $F_0$ changes. Japanese is a ''mora-timed'' pitch-accent language: that means there is an accent nucleus position counted in mora units within an accentual phrase. Pitch accents have a large impact on the perceptual naturalness of speech because incorrect pitch accents may be judged as incorrect ``pronunciations`` by listeners even if they have correct phone realization. Moreover, accentual phrases in Japanese normally have mora of varying lengths. Since the length of an accentual phrase could be very long, we hypothesize that long-term information plays a significantly important role in TTS for pitch accent languages. 

Therefore, we propose a modified architecture by introducing ''self-attention'' after LSTM layers at the encoder and decoder as illustrated in Figure \ref{fig:tacotron}-B. It is known that by directly connecting distant states, self-attention relieves the high burden placed on LSTM to learn long-term dependencies to sequentially propagate information over long distances \cite{Lin2017}. This extension is inspired from a sequence-to-sequence neural machine translation architecture proposed by \cite{Vaswani2017}. We refer to this architecture as \textit{SA-Tacotron}.

The self-attention block consists of self-attention, followed by a fully connected layer with tanh activation and residual connection. We use multi-head dot product attention \cite{Vaswani2017} as an implementation of self-attention. This block is inserted after LSTM layers at the encoder and decoder.
At the encoder, the output of CBH-LSTM layers is processed with the self-attention block. Since LSTM can capture the sequential relationships of inputs, we do not use positional encoding \cite{DBLP:journals/corr/abs-1809-08895}. Both self-attended representation and the original output of the CBH-LSTM layers are final outputs of the encoder. 

At the decoder, the two outputs from the encoder are attended with a dual source attention mechanism \cite{BLP:journals/corr/ZophK16}. We choose a different attention mechanism for each source, forward attention for the output of CBH-LSTM and additive attention for the self-attended values. This is because we want to utilize the benefits of both: forward attention accelerates alignment construction, and additive attention provides flexibility to select long-term information from any segment. In addition, we can visualize both alignments.
Unlike the encoder, self-attention works autoregressively at the decoder. At each time step of decoding, the self-attention layer attends all past frames of LSTM outputs and outputs only the latest frames as a prediction output. The predicted frames are fed back as input for the next time step.\footnote{At training time, since all target frames are available, this computation can be parallelized by applying a step mask. Since the decoder depends on LSTM, the whole computation cannot be parallelized, but this optimization decreases memory consumption because all past LSTM outputs do not need to be preserved at each time step to calculate gradients on a backward path in backpropagation algorithm. Thanks to this optimization, we can train the extended architecture with a negligible increase in training time.}

\vspace{-2mm}
\subsection{Tacotron using vocoder parameters}
\label{subsec:tacotron_vocoder_parameter}

Explicitly modeling the fundamental frequency ($F_0$) might be a more appropriate choice for TTS systems for pitch-accent languages. To incorporate $F_0$ into the proposed systems, we further developed a variant of \textit{SA-Tacotron} by using vocoder parameters as targets. We use mel-generalized cepstrum coefficients (MGC) and discretized $\log F_0$ as vocoder parameters, and we predict these parameters with Tacotron. We choose 5~ms for the frame shift to extract MGC and $F_0$ as such fine-grained analysis conditions are typically required for reliable speech analysis based on vocoderes.
However, note that this condition is not a natural choice for training Tacotron, which typically uses coarse-grained condition, usually 12.5~ms frame shifts and 50~ms frame lengths, to reduce input and output mismatch. With a frame shift of 5~ms, the length of target vocoder parameter sequences becomes 2.5 times longer than the normal 12.5~ms condition. In other words 2.5 times longer autoregressive loop iteration is required to predict a target, so this task is much more challenging. To alleviate the difficulty, we set the reduction factor to be three in order to reduce the target length. This setting results in 5/3 times longer target length compared to \textit{SA-Tacotron} in the previous section.\footnote{We tried larger reduction factors, but the audio quality deteriorated as the reduction factor increased.}

Figure~\ref{fig:tacotron}-C shows the modified architecture of the \textit{SA-Tacotron} using MGC and $\log F_0$ as targets. To handle the two types of vocoder parameters, we introduce two pre-nets and three output layers at the decoder. The output layers include a MGC prediction layer that consists of two fully connected layers followed by tanh and linear activations, a $\log F_0$ prediction layer which is a fully connected layer followed by softmax activation, and a stop flag prediction layer, which is a fully connected layer followed by sigmoid activation. We represented discretized $\log F_0$ as one-hot labels at training time, but feed back predicted probability values at inference time \cite{wangxinDARf0}. We use L1 loss for MGC and cross entropy error for discretized $\log F_0$ and stop flag, and we optimize the model by using the weighted sum of the three losses. The cross entropy error of $\log F_0$ is scaled by 0.45 to adjust its order to the other two loss terms.


\section{Experiments}
\vspace{-2mm}
\subsection{Experimental conditions}

We used a Japanese speech corpus from the ATR Ximera dataset \cite{kawai2006ximera}. This corpus contains 28,959 utterances from a female speaker and is around 46.9 hours in duration. 
The linguistic features, such as phoneme and accentual-type label, were manually annotated, and the phoneme label had 58 classes, including silence, pause, and short pause \cite{Luong2018}.
To train our proposed systems, we trimmed the beginning and ending silence from the utterances, after which the duration of the corpus was reduced to 33.5 hours. We used 27,999 utterances for training, 480 for validation, and 142 for testing.

For the experiment, we built several TTS systems as listed in Table~\ref{tbl:systems}. The \textit{JA-Tacotron} and \textit{SA-Tacotron} with and without accentual-type labels were built to show whether the investigated architectures can learn lexical pitch accents in an unsupervised manner. 
We also built a \textit{SA-Tacotron} that uses vocoder parameters instead of mel-spectrogram as the acoustic features. 
In addition, we included \textit{JA-Tacotron} with forced alignment instead of predicted alignment to understand the accuracy of duration modeling better. With forced alignment, alignments are calculated with teacher forcing, and target acoustic parameters are predicted with the alignments obtained with teacher forcing. Note that, in this setting, even though forced alignments are calculated with teacher forcing, acoustic parameter prediction itself does not use teacher forcing. 
%
%
%

For \textit{JA-Tacotron} and \textit{SA-Tacotron}, we allocated 32 dimensions for accentual-type embedding and 224 dimensions for phoneme embedding. For the models without accentual-type embedding, 256 dimensions were allocated to phoneme embedding. We set the reduction factor to be two for the models using mel-spectrogram as a target and three for the models using vocoder parameters. All the predicted frames of the acoustic features were fed back as the next input. At inference time, the inference was stopped on the basis of a binary stop flag as in \cite{Shen2017}. The network was optimized with Adam optimizer \cite{DBLP:journals/corr/KingmaB14}. We used exponential learning decay with an initial rate 0.0005 for the models using mel-spectrogram, and 0.002 for the models using vocoder parameters. We implemented our proposed systems using TensorFlow%
\footnote{The source codes is availabe at https://github.com/nii-yamagishilab/self-attention-tacotron}.

For baseline systems, we included two classical pipeline systems that use vocoder parameters and mel-spectrogram \cite{Luong2018}, \cite{Lorenzo-Trueba2018}, \cite{XinWang2018}. Unlike the architecture of our proposed systems, these pipeline systems used full context labels as linguistic features and needed to have duration prediction models. To test how the accuracy of duration prediction affects the naturalness of synthetic speech, we compared phone duration predicted by a hidden semi-Markov model (HSMM) with oracle alignments obtained by force alignments. 
Finally, as a reference for how much listeners are sensitive to incorrect lexical pitch accents, a baseline with slightly corrupted accentual labels was also included.\footnote{This system is named MOC in \cite{Luong2018}.}

Two types of WaveNet models were trained for the experiment, one taking the mel-spectrograms as the input and the other using the MGC and $F_0$ (vocoder parameters). These two WaveNets had the same network structure as that in our previous study \cite{XinWang2018}.

\vspace{-2mm}
\subsection{Objective evaluation}

\begin{figure}[!t]
    \begin{center}
    \begin{subfigure}[t]{0.43\textwidth}
    {\includegraphics[width=\textwidth]{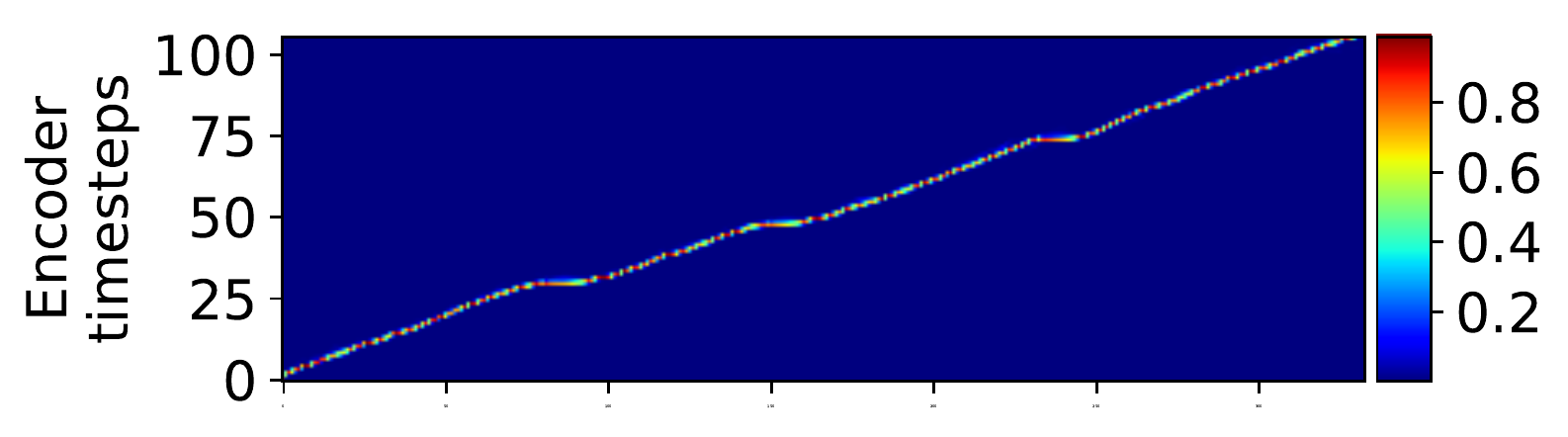}}
    \end{subfigure}
    \vspace{-3mm}
    \begin{subfigure}[t]{0.43\textwidth}
    {\includegraphics[width=\textwidth]{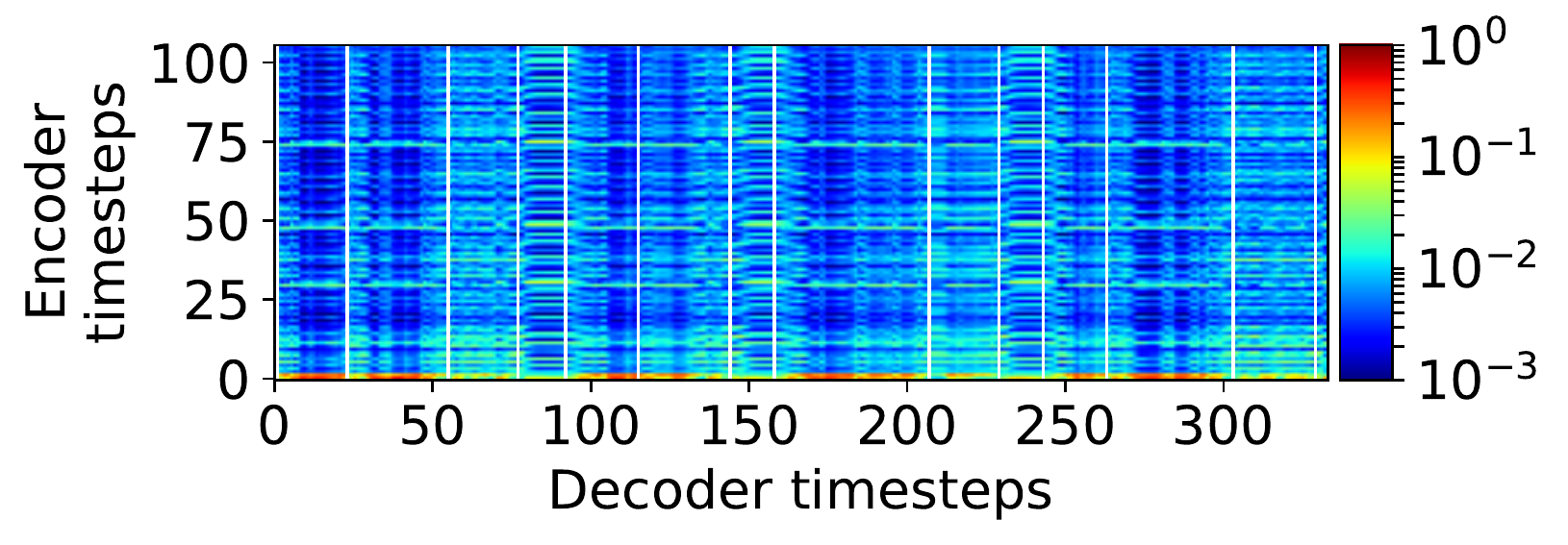}}
    \end{subfigure}
    \end{center}
    \vspace{-5mm}
    \caption{Alignment obtained by dual source attention in \texttt{SATMAP}. Top figure shows alignment between output of encoder's LSTM layer and target mel-spectrogram (forward attention). Bottom figure shows alignment between output of encoder's self-attention block and target mel-spectrogram (additive attention). Vertical white lines indicate accentual phrase boundaries obtained by forward attention.}
    \label{fig:self-attention-visualization}
    \vspace{-4mm}
\end{figure}

\noindent 
\textbf{What does self-attention learn?:}
\label{subsec:alignment-visualization}
Figure \ref{fig:self-attention-visualization} shows a visualization of the attention layers of \textit{SA-Tacotron} learned on the Japanese corpus. The first figure from the top shows the alignment of an encoder LSTM source and mel-spectrogram target for dual source attention. We can clearly see a sharp monotonic alignment formed by the forward attention. The second figure from the top shows the alignment of an encoder self-attention source and mel-spectrogram target. It seems to be related to accentual phrase segments and phrase breaks divided by pauses.


\begin{figure}[tb]
\begin{center}
\includegraphics[width=0.75\columnwidth]{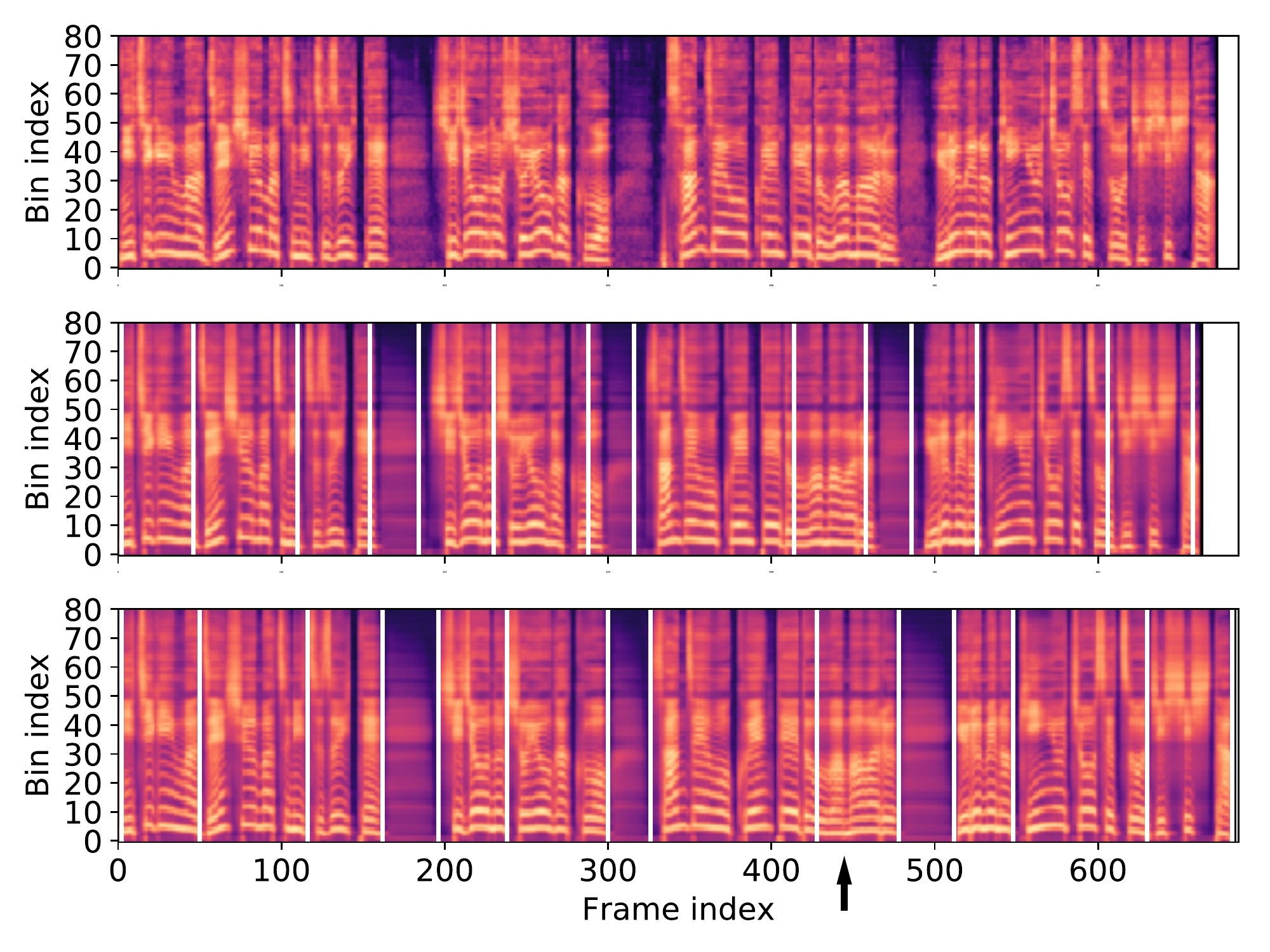}
\vspace{-20pt}     
\end{center}
\caption{Natural mel-spectrogram (top figure), mel-spectrogram predicted from \textit{SA-Tacotron} with accentual-type labels (middle figure), and mel-spectrogram predicted from \textit{SA-Tacotron} without labels (bottom figure). Black arrow in the bottom figure points wrong harmonics that results in wrong accent. White lines show accentual phrase boundaries acquired from attention's output.}
\label{fig:mel-spectrogram-comparison}
\vspace{-10pt}   
\end{figure}

\noindent 
\textbf{What is the effect of accentual-type labels?:}
\label{subsec:visual_comparison}
Figure \ref{fig:mel-spectrogram-comparison} shows predicted mel-spectrograms from \textit{SA-Tacotron} with and without accentual-type labels. Accentual phrase boundaries predicted by the attention mechanism are also shown in the figure. From this figure, through comparison with a natural spectrogram, we see that the predicted spectrogram from \textit{SA-Tacotron} without labels has wrong accentual positions and harmonics, whereas that from \textit{SA-Tacotron} with labels does not. 
From informal listening, we also noticed that \textit{SA-Tacotron} without labels had incorrect accent nucleus positions.

\noindent 
\textbf{Comparison of mel-spectrogram and vocoder parameters:}

The alignment between source phoneme and target spectrogram frames should monotonically increase. Non-monotonic alignment may result in mispronunciation, some phonemes being skipped, repetition, the same phoneme continuing, and intermediate termination. We therefore manually counted abnormal alignment errors included in the test set. We observed no alignment errors for \textit{JA-Tacotron} and \textit{SA-Tacotron} using mel-spectrograms as a target. However, alignment errors were found for \textit{SA-Tacotron} using vocoder parameters due to the longer length than the corresponding mel-spectrogram. We found 19 alignment errors out of 142 test utterances.


\vspace{-2mm}
\subsection{Subjective evaluation}
\label{subsec:listening_test}

\begin{figure}[tb]
\begin{center}
\vspace{-14pt}   
\includegraphics[width=0.85\columnwidth]{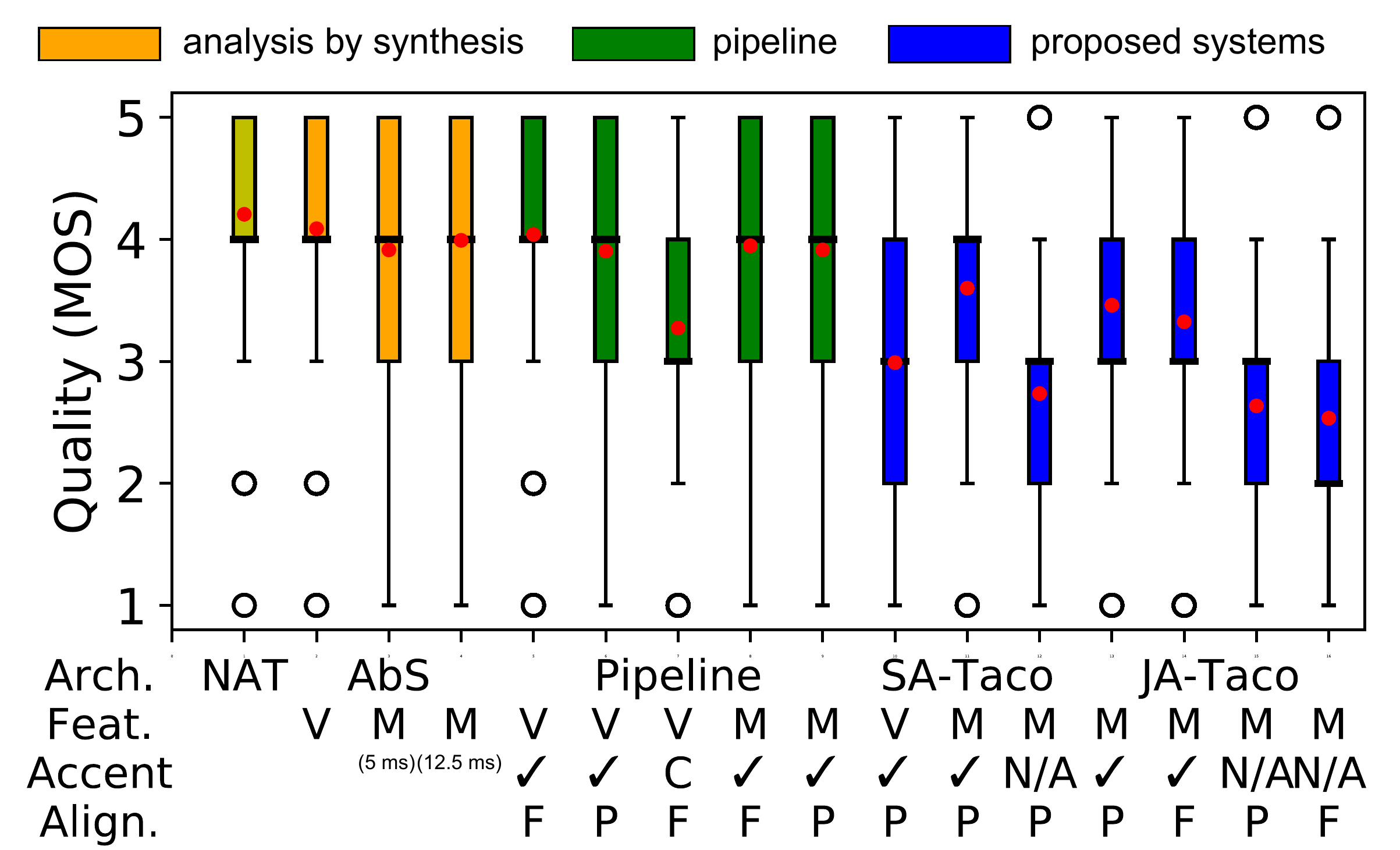}
\vspace{-16pt}     
\end{center}
\caption{Box plots of MOS scores of each system regarding naturalness of synthetic speech. Red circles represent average values. NAT indicates natural speech. Refer to Table \ref{tbl:systems} for notations.}
\label{fig:mos}
\vspace{-10pt}   
\end{figure}

We recruited 236 native Japanese speakers as listeners by crowdsourcing. The listeners evaluated 32 samples from 16 systems in a single test set. This includes natural speech and analysis by synthesis (copy synthesis). One listener can evaluated at most 10 test sets. One sample was evaluated 20 times and we got 45,440 data points in total. Figure \ref{fig:mos} shows five-point mean opinion scores of the proposed and baseline systems for the listening test results. Statistical significance was analyzed using the two-sided Mann-Whitney statistical test.

\begin{table}[tb]
\caption{TTS systems used for our analysis. Notations are \texttt{V}: vocoder parameters, \texttt{M}: mel spectrogram, \texttt{A}: accentual type label, \texttt{N}: no accentual type label, \texttt{P}: predicted alignment, \texttt{F}: forced alignment.}
\label{tbl:systems}
\begin{center}
\scriptsize
\vspace{-20pt}   
\begin{tabular}{|c|c|c|c|c|c|}\hline
System & Architecture & Acoustic feature & Accent label & Alignment \\\hline\hline
\texttt{SATVAP} & \multirow{3}{*}{\textit{SA-Tacotron}} & MGC \& $F_0$ & \checkmark & predicted \\
\texttt{SATMAP} & & Mel-spec. 12.5~ms & \checkmark & predicted \\
\texttt{SATMNP} & & Mel-spec. 12.5~ms & N/A & predicted \\
\hline
\texttt{TACMAP} & \multirow{4}{*}{\textit{JA-Tacotron}} & \multirow{4}{*}{{Mel-spec. 12.5~ms}} & \checkmark & predicted \\
\texttt{TACMAF} & 							    & 						 & \checkmark & force-aligned \\
\texttt{TACMNP} & & & N/A & predicted \\
\texttt{TACMNF} & & & N/A & force-aligned \\
\hline
\texttt{PIPVAF} & \multirow{5}{1cm}{Pipeline \cite{Luong2018,XinWang2018}} & MGC \& $F_0$ & \checkmark & force-aligned \\
\texttt{PIPVAP} & & MGC \& $F_0$ & \checkmark& predicted \\
\texttt{PIPVCF} & & MGC \& $F_0$ & corrupted & force-aligned \\
\texttt{PIPMAF} & & Mel-spec. 5~ms & \checkmark & force-aligned\\
\texttt{PIPMAP} & & Mel-spec. 5~ms & \checkmark & predicted \\
\hline
\end{tabular}
\vspace{-24pt}   
\end{center}
\end{table}

\noindent 
\textbf{What is the effect of accentual-type labels?:}
All proposed systems without accentual-type labels got significantly lower scores than the corresponding systems with labels; for example, JA-Tacotron without labels had a score of $2.63 \pm 0.03$ whereas \textit{JA-Tacotron} with labels got $3.46 \pm 0.03$. This means that the architectures of the proposed systems cannot learn lexical pitch accents in an unsupervised fashion and require additional inputs. 
The pipeline system with corrupted labels also showed a significant drop with a score of $3.27 \pm 0.03$. This shows that incorrect accents affected listener's judgments towards the naturalness of the synthetic speech.

\noindent 
\textbf{Does self-attention help?:}
\textit{SA-Tacotron} had better scores than \textit{JA-Tacotron} for each condition with or without accentual-type labels. This indicates that self-attention layers have a positive effect on the naturalness. Among our proposed systems, \textit{SA-Tacotron} with labels (\texttt{SATMAP}) got the highest score of $3.60 \pm 0.03$. 

\noindent 
\textbf{Comparison of mel-spectrogram and vocoder parameters:}
\textit{SA-Tacotron} using vocoder parameters got a relatively low score, $2.99 \pm 0.03$, even if it used accentual-type labels and self-attention layers. This is because this system generated alignment errors due to the prediction of longer sequences as we described in the previous section. Among the baseline systems, the systems using MGC and $F_0$ had higher scores than the systems using mel-spectrogram under both the forced and predicted alignment conditions.

\noindent 
\textbf{Comparison of predicted and forced alignment:}
Interestingly, \textit{JA-Tacotron} using forced alignment got lower scores than that using predicted alignment under both conditions with and without accentual-type labels. This result is surprising because, in traditional pipelines, forced alignment is used as an oracle alignment and normally leads to better perceptual quality than that of the predicted case. Since Tacotron learns both spectrograms and alignments simultaneously, it seems to produce the best spectrograms when it infers both of them. Among the baseline pipeline systems, as expected, a forced alignment gave higher scores than predicted alignment for both systems using vocoder parameters and mel-spectrogram. In the case of predicted alignment, the score has a long tail variance towards the low score region. 

\noindent 
\textbf{Comparison of pipeline and Tacotron systems:}
The best proposed system still does not match the quality of the best pipeline system. 
\textit{SA-Tacotron} with accentual-type labels and the pipeline system using mel-spectrogram and predicted alignment had $3.60 \pm 0.03$ and $3.90 \pm 0.03$, respectively. These are not the same results as for the English experiments reported in \cite{Shen2017}.
One major difference of our proposed systems from pipeline systems other than architecture is input linguistic features; our proposed systems use phoneme and accentual-type labels only, but the baseline pipeline systems use various linguistic labels including word-level information such as inflected forms, conjugation types, and part-of-speech tags. 
In particular, an investigation on the same Japanese corpus found that the conjugation type of the next word is quite useful for $F_0$ prediction \cite{wangPhD}.


\vspace{-5mm}
\section{Conclusion}
\label{sec:con}

In this paper, we applied Tacotron to Japanese to extend it to a pitch-accent language. We proposed phone-based Tacotrons with and without accentual-type labels, one with self-attention layers to capture long term information better, and one using vocoder parameters including fundamental frequency. We conducted objective and subjective evaluations. Among the proposed systems, Tacotron with the self-attention extension outperformed that without self-attention both with and without labels. 
However, we revealed that, unlike experiments reported for English, the quality of traditional pipeline systems is better than the proposed systems for Japanese. We also found that choosing vocoder parameters is beneficial to pipeline systems, but this is completely opposite for the case of Tacotron. 

One major difference of our proposed systems from the pipeline systems is the absence of word level information in linguistic features, so incorporating this information may improve the quality of the proposed systems and bring them up to the pipeline system's level. Our next step towards end-to-end speech synthesis in various languages is to incorporate word-level information such as Kanji. 


\vspace{0.5mm}
\noindent 
\textbf{Acknowledgements}
We are grateful to Prof.\ Zhen-Hua Ling from USTC for kindly answering our questions.

\vfill\pagebreak

\bibliographystyle{IEEEbib}
\bibliography{BIB}

\newpage 

\appendix 

\section{Hyper-parameters}
Table~\ref{tbl:hyperparameters} shows the hyper-parameters used for \textit{SA-Tacotron} with accentual-type embedding. 

\begin{table}[h]
\caption{Hyper-parameters}
\label{tbl:hyperparameters}
\begin{center}
\vspace{-12pt}   
\scriptsize
\begin{tabular}{|l|l|}\hline
Frame length, shift & 50 ms, 12.5 ms\\\hline
Sample rate, FFT size & 48 kHz, 4096\\\hline
Embeddings & Phoneme: 224-D, Accent: 32-D\\\hline
Encoder pre-net & \scriptsize{Phoneme: 224/112-D, Accent: 32/16-D}\\\hline
Attention RNN & 256-D cells, 10-D kernels, 5-D filters\\\hline
Encoder \& decoder LSTM & 256-D cells, 10~\% zoneout rate\\\hline
Encoder self-attention & 32-D, 2 heads, 1 hop, 5~\% drop rate\\\hline
Decoder self-attention & 256-D, 2 heads, 1 hop, 5~\% drop rate\\\hline
\end{tabular}
\vspace{-15pt}   
\end{center}
\end{table}

\section{Objective evaluation of predicted F0}
\label{subsec:objective_evaluation}

Furthermore, we conducted an objective evaluation for four variant systems by using vocoder parameters such as JA-Tacotron and SA-Tacotron with and without accentual-type label.

To evaluate the $F_0$ prediction capability of JA-Tacotron and SA-Tacotron, we evaluated the objective metrics of $F_0$. To calculate the metrics we adjusted the frames between the predicted and ground truth $F_0$ by using force alignment. JA-Tacotron using vocoder parameters without labels failed to learn alignments between the source and target, so we did not include it.

\begin{table}[h]
\caption{Objective evaluation of $F_0$ predicted by JA-Tacotron and SA-Tacotron.}
\label{tbl:f0_objective_metrics}
\begin{center}
\scriptsize
\vspace{-12pt}   
\begin{tabular}{|c|c|c|c|c|}\hline
System & accent & RMSE & CORR & U/V\\\hline
\texttt{TACVAF} & \checkmark & 31.81 & 0.88 & 7.10~\%\\\hline
\texttt{SATVAF} & \checkmark  & 31.77 & 0.88 & 7.20~\% \\\hline
\texttt{SATVNF} & N/A & 39.30 & 0.79 & 7.04~\% \\\hline\hline
\texttt{PIPVAF} & \checkmark & 23.31 & 0.94 & 3.25~\% \\\hline
\texttt{PIPVCF} & corrupted & 31.09 & 0.89 & 3.29~\% \\\hline
\end{tabular}
\vspace{-15pt}   
\end{center}
\end{table}

Table~\ref{tbl:f0_objective_metrics} shows RMSE, correlation and U/V errors of $F_0$. Both JA-Tacotron and SA-Tacotron with accentual-type labels had an $F_0$ correlation value of 0.88. This indicates that self-attention had no effect on $F_0$ prediction accuracy. The systems without labels show lower correlation of 0.79 compared to the systems with labels, because of wrong accents as described in Section \ref{subsec:visual_comparison}. Although these values were still lower than a baseline pipeline system that had 0.94 \cite{Luong2018}, we think they are good enough considering that a frame shift of 5~ms is not the best condition for Tacotron. In addition, forced alignment itself has a negative effect on audio quality in Tacotron as we described in Section \ref{subsec:listening_test}. The baseline system with noisy accentual-type labels had a correlation of 0.89. The noisy baseline had artificial accent errors with a probability of 50~\%. Even though this is almost same as the correlation values of our proposed systems with accentual-type label, we do not think our proposed systems has accent errors with a probability of 50~\%. As can be seen in the listening test result in Section \ref{subsec:listening_test}, our proposed systems using mel-spectrogram with labels outperform the noisy baseline, so the relatively low correlation of $F_0$ was caused by the unsuitable conditions of acoustic features for Tacotron.

\section{Visual and statistical analysis of self-attention at encoder and decoder}
\label{subsec:self-attention heads visualization}

Because alignments of self-attention at encoder in \textit{SA-Tacotron} are hard to interpret at sample level, we conducted statistical analysis for alignment scores of self-attention in a test sets. We calculate mean alignment scores of phoneme pairs that frequently occur (more than 30 times). In head 1 we found some alignments based on similarity. For example, top three phonemes with high score values are identical phoneme pairs and there are 11 identical phoneme pairs within top 100. In addition, we found 13 phoneme pairs that belong to same group (e.g. long vowels) within top 100. The head 2 showed strong affinity to silence and pauses; we found 88 pairs that include pause or silence out of top 100 pairs with high alignment scores.

\begin{figure}[tb]
\begin{tabular}{c}
\includegraphics[width=0.51\columnwidth]{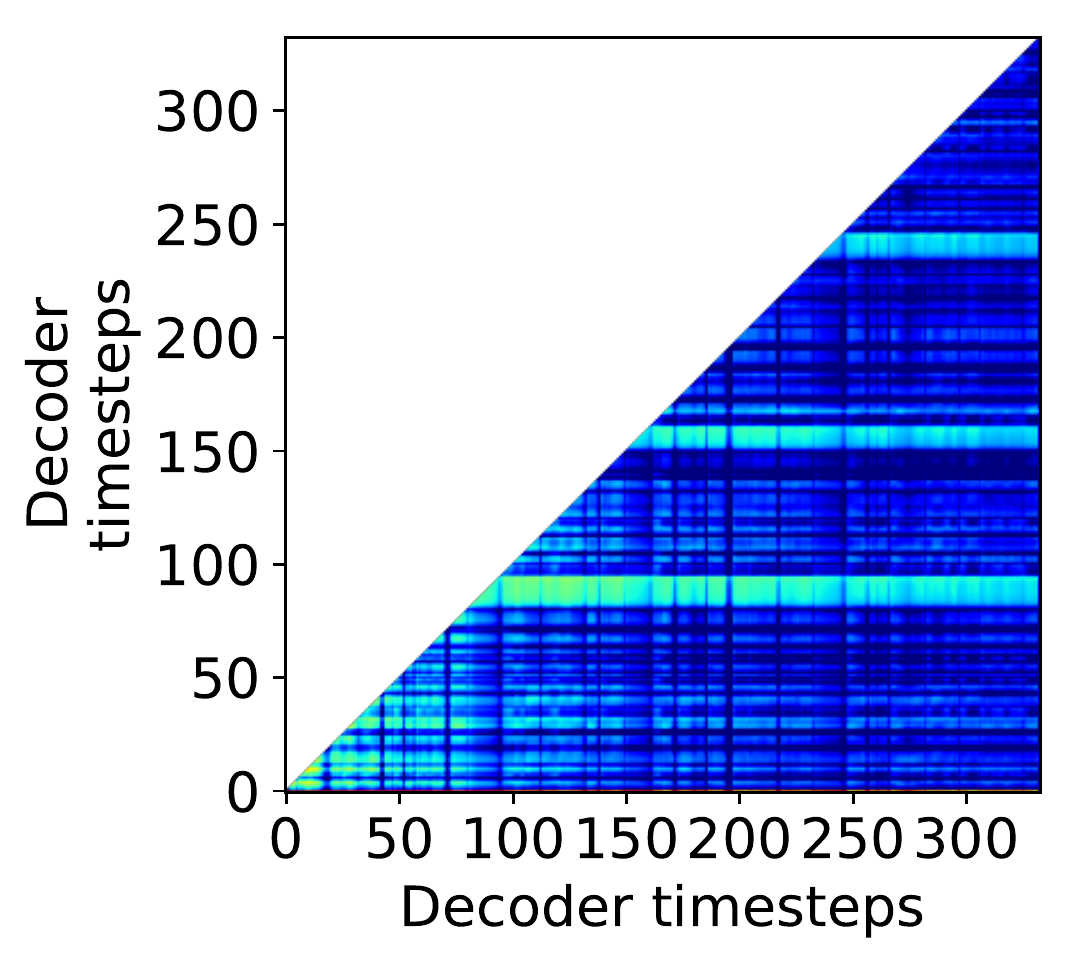}
\includegraphics[width=0.49\columnwidth]{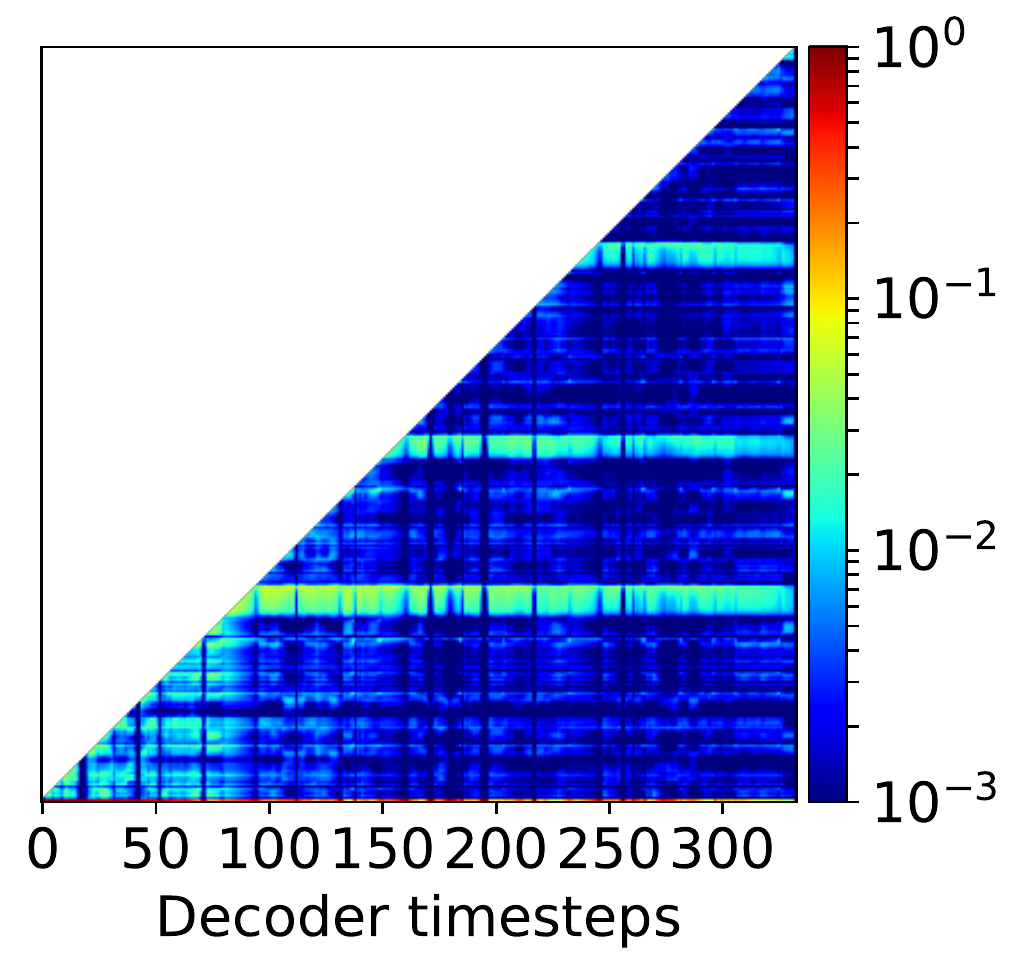}
\vspace{-15pt}     
\end{tabular}
\caption{Alignment visualization of two heads from self-attention layer at decoder in \textit{SA-Tacotron}. Three horizontal bands with relatively high activation correspond to pause positions.}
\label{fig:decoder-self-attention}
\vspace{-20pt}   
\end{figure}

Fig.~\ref{fig:decoder-self-attention} shows alignments of two heads from decoder self-attention in \textit{SA-Tacotron}. Note that the upper triangular part above diagonal cannot be attended because it is the future information. Although both heads of the self-attention layer attend mostly the first frame which is silence with no useful information, they weekly attends almost broad range of past frames. Both heads attend all past pauses once pauses are encountered.

\end{document}